\documentclass[a4paper]{article}

\usepackage{INTERSPEECH2021}
\usepackage{units}
\graphicspath{{./fig/}}

\title{A DNN Based Post-Filter to Enhance the Quality of Coded Speech in MDCT Domain}
\name{Kishan Gupta$^1$,  Srikanth Korse$^1$,  Bernd Edler$^{1,2}$, Guillaume Fuchs$^1$}
\address{
  $^1$ Fraunhofer IIS, Erlangen, Germany. 
  $^2$ International Audio Laboratories, Friedrich-Alexander University (FAU), Erlangen-Nürnberg, Germany$^*$
   \thanks{$^*$ International Audio Laboratories is a joint institution between Fraunhofer IIS and Friedrich-Alexander University (FAU)}   
	} 
\email{kishan.gupta@iis.fraunhofer.de}

\begin{document}

\maketitle
\begin{abstract}
Frequency domain processing, and in particular the use of Modified Discrete Cosine Transform (MDCT), is the most widespread approach to audio coding. However, at low bitrates, audio quality, especially for speech, degrades drastically due to the lack of available bits to directly code the transform coefficients. Traditionally, post-filtering has been used to mitigate artefacts in the coded speech by exploiting a-priori information of the source and extra transmitted parameters. Recently, data-driven post-filters have shown better results, but at the cost of significant additional complexity and delay. In this work, we  propose a mask-based post-filter operating directly in MDCT domain of the codec, inducing no extra delay.  The real-valued mask is applied to the quantized MDCT coefficients and is estimated from a relatively lightweight convolutional encoder-decoder network. Our solution is tested on the recently standardized low-delay, low-complexity codec (LC3) at lowest possible bitrate of 16 kbps.
Objective and subjective assessments clearly show the advantage of this approach over the conventional post-filter, with an average improvement of 10 MUSHRA points over the LC3 coded speech. 
\end{abstract}
\vspace{0.02\linewidth}

\section{Introduction} \label{sec:Introduction}
\paragraph*{}
State-of-the-art speech and audio communication codecs such as 3GPP Extended Adaptive Multi-Rate-Wideband (AMR-WB+)~\cite{AMRWB+} and 3GPP Enhanced Voice Services (EVS)~\cite{EVS:2014alt} typically use Code-Excited Linear Prediction (CELP) and transform coding to encode speech and music, respectively, at lower bitrates. However, CELP-based coding has higher complexity compared to transform coding, especially at the encoder side. Therefore, the recently standardized low complexity, low delay codec (LC3)~\cite{LC3:2018Std}~\cite{schnell2021lc3} completely relies on transform coding which involves quantizing and coding the spectral coefficients after an MDCT (Modified Discrete Cosine Transform), thus reducing the complexity by a factor of 6 compared to EVS~\cite{EVS:2014alt} in super-wideband mode. At medium to higher bitrates, due to the availability of sufficient bits, transform-based coding yields sufficiently good to transparent quality. Conversely, at low bitrates, due to insufficient bits, spectral holes are created, leading to audible artefacts~\cite{disch2016intelligent}.

To enhance the perceptual quality of coded speech at these low bitrates, tools such as noise filling, gap filling~\cite{disch2016intelligent} and LTPF (Long Term Post-filter) are employed~\cite{EVS:2014alt}~\cite{Fuchs_2015}. While noise filling and gap filling typically aid in mitigating the audible artefacts by treating the spectral holes, the LTPF tries to improve the voiced parts of coded speech by the attenuating inter-harmonic noise~\cite{chen1995adaptive}. All of the above-mentioned techniques require the transmission of additional information to the decoder as side information, hence causing an overhead in the bit consumption.

In recent years, several data-driven post-filters which solely rely on the statistics obtained from the coded speech have been proposed in order to enhance the quality of coded speech~\cite{Das2020_ESSV}~\cite{Korse_2020}~\cite{Biswas2020}~\cite{Skoglund2019}. While~\cite{Das2020_ESSV} designs a post-filter in the MDCT domain, based on a simplistic statistical model of the quantization noise, ~\cite{Korse_2020} trains a DNN to estimate a real-valued mask per time-frequency tile based on log-magnitude as input in the STFT (Short Time Fourier Transform) domain. In contrast, both~\cite{Biswas2020} and~\cite{Skoglund2019} have proposed a post-filter in the time-domain using generative models such as GAN (Generative Adversarial Networks) and LPCNet, respectively. While post-filters based on generative models have the possibility of processing both magnitude and phase in contrast to methods that operate only on magnitude, they suffer from a significant complexity overhead and can be prone to lack of generalization for unseen speakers. In addition,~\cite{Skoglund2019} relies on spectral features from decoded speech, and also needs features derived from LPC coefficients in bitstream which are usually unavailable in transform coding whereas~\cite{Korse_2020} needs to perform a forward and inverse STFT transform for the enhancement. 

In this paper, we propose a mask-based post-filter that operates in the MDCT domain. Instead of working on decoded speech signal, our proposed post-filter can directly enhance the quantized MDCT coefficient available at LC3 decoder before inverse transformation, thus saving overhead caused by an additional analysis, synthesis or feature extraction. We discuss the constraint associated with mask-based approach in MDCT domain as it has been shown that a simple ratio mask-based approach similar to~\cite{Korse_2020} when directly applied to MDCT coefficients produces audible artefacts~\cite{Kuech_2007}~\cite{Koizumi2018}. To mitigate such artefacts, we propose to train our model to estimate a magnitude mask from the MCLT (Modulated Complex Lapped Transform) domain and show that such mask can be used to enhance MDCT coefficients during inference. We also show that such a training method does not require an inverse transform during DNN training and avoids the need to compute the loss in time-domain as suggested in~\cite{Koizumi2018}.


\section{Problem Formulation}

\subsection{System Overview} 
\begin{figure}[t]
	\centering
	\includegraphics[width=\linewidth]{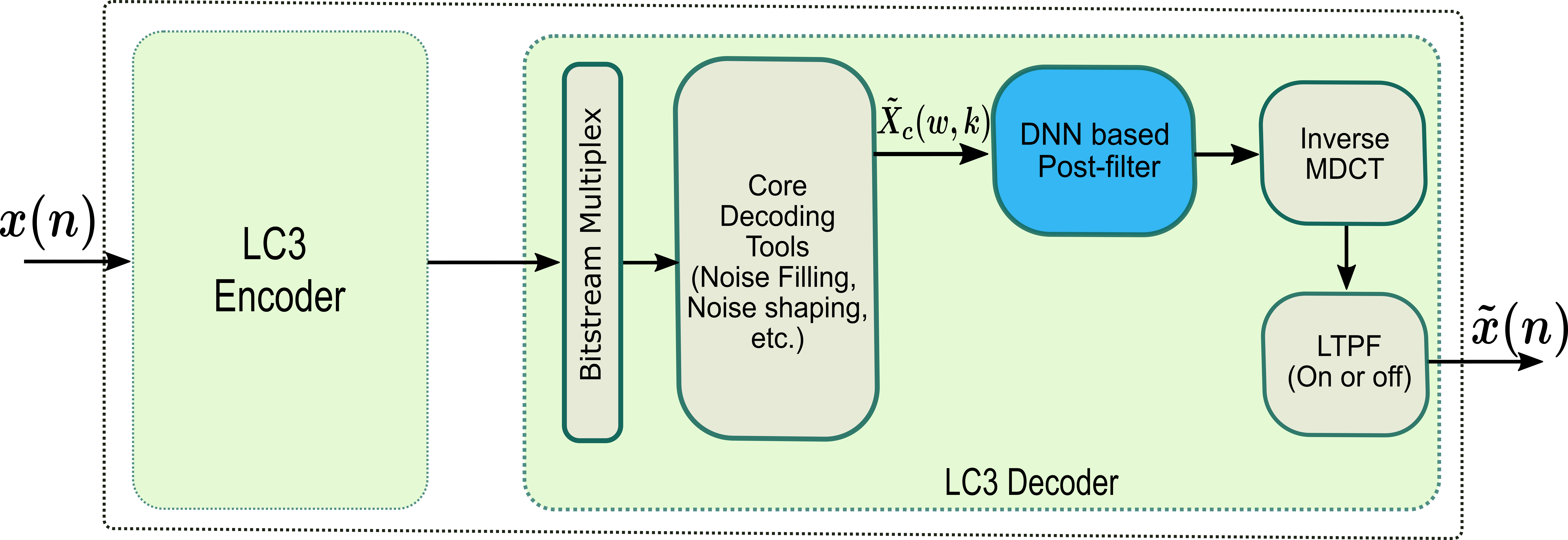}
	\caption{System overview of the proposed DNN based post-filter}
	\label{fig:reference_system}
\end{figure}

Fig.~\ref{fig:reference_system} shows the integration of our proposed post-filter with the MDCT-based LC3 codec. In such a setup, the post-filter operates in the MDCT domain at the decoder side, before the inverse transformation into time domain. It does not require additional feature extraction or time-frequency analysis, but is constrained by the MDCT transform used in the codec. In our experimental setup, we use LC3 with 10~\unit{ms} frames~\cite{schnell2021lc3}, which is then inherited by the post-filter. 

\subsection{Mask Formulation}
In simple terms, coded speech can be described as:
\begin{equation}
	\label{equ_quantNoise}
	\tilde{x}(n) = x(n) + \delta(n),
\end{equation} 

\noindent where $x(n)$ is uncoded speech and $\delta(n)$ is the quantization noise. In a transform codec, the quantization noise is the approximation error arising from the spectral quantization. Spectral noise shaping based on a perceptual model is used to make the quantization noise less perceivable. As a result, the introduced quantization noise is correlated to the speech signal. 

A post-filter that predicts a real-valued mask used on real-valued transform coefficients (e.g. MDCT coefficients) can be used to clean the quantization noise resulting from transform coding. However, the MDCT domain is not well suited for signal manipulation in the frequency domain for several reasons~\cite{wangmdct}. Its basis vectors are not shift-invariant, and MDCT does not conserve energy. A perfect reconstruction can only be done by considering adjacent windows and the principle of time domain aliasing cancellation (TDAC). Any manipulation in the MDCT domain can affect these conditions and impact resulting time aliasing~\cite{Kuech_2007}. Moreover, the MDCT coefficients are real-valued and cannot be easily interpreted in terms of magnitude and phase. Therefore, we propose to train our model to estimate the real-valued magnitude mask computed on magnitude spectrum of the MCLT, a complex-valued transform similar to STFT but with time and frequency shifts, for which the MDCT is given by its real part.

The MCLT of the time-domain signal $x(n)$is given by:

\begin{equation}
	\label{equ_MCLT}
	X(w,k) = X_C(w,k) + j X_S(w,k),
\end{equation} 

\noindent where $w$ and $k$ are the time and frequency index of the MCLT bins, respectively,  $X_C(w,k)$ are the MDCT and $X_S(w,k)$ are the MDST (Modified Discrete Sine Transform) of the time-domain signal and are defined as:

\begin{equation}
	\label{equ_MDCT}
	X_C(w,k) = \sum_{n = 0}^{2N-1}h(n)x(n)\cos\left[\frac{\pi}{N} (n + \frac{1}{2} + \frac{N}{2})(k + \frac{1}{2})\right],
\end{equation} 

\begin{equation}
	\label{equ_MDST}
	X_S(w,k) = \sum_{n = 0}^{2N-1}h(n)x(n)\sin\left[\frac{\pi}{N} (n + \frac{1}{2} + \frac{N}{2})(k + \frac{1}{2})\right],
\end{equation} 

\noindent where $h(n)$ is a low delay asymmetric window used in LC3~\cite{schnell2021lc3}, $x(n)$ is the input signal of length $2N$ and $ k = 0, 1,....,N-1$. The MCLT maps $2N$ input samples to $N$ complex output coefficients. It is then straightforward to design an optimal filter by considering the magnitude of the complex-transform. We define the ideal magnitude mask of our post-filter in MCLT domain as,

\begin{equation}
	\label{equ_IRM}
	M(w, k) = \frac{|X(w,k)|}{|\tilde{X}(w,k)|+\gamma},
\end{equation}

\noindent where $|X(w,k)|$ and $|\tilde{X}(w,k)|$ denote the MCLT magnitude of clean and coded speech, respectively. A small constant $\gamma$ is added to prevent division by zero. The so-obtained magnitude mask can be applied to the MDCT coefficients ignoring the MDST components during the inverse transform, which results in the following processed MDCT coefficients:

\begin{equation}\label{equ_Mask_mdct}
	\hat{X}_C(w,k) = M(w,k) \cdot \tilde{X}_C(w,k),
\end{equation}

The MDCT does not explicitly carry phase information, but also not the exact magnitude information. The processing of the MDCT coefficients in Eq.~\eqref{equ_Mask_mdct} with a mask derived from the MCLT magnitude spectrum is able to simulate a magnitude manipulation in the MDCT domain. It can be then assumed that the phase is either unaffected or only very slightly affected by the so-derived masking operation. The post-filter usage is then greatly simplified, and no specific care is required to avoid artefacts arising from time-domain aliasing caused when the TDAC principle is broken by manipulating the MDCT coefficients.

In our proposed post-filter, the model takes MDCT alone as input and predict an optimal mask computed on the MCLT. The ability of a DNN to achieve such a prediction is not overly surprising since the spectrum of MDCT and MCLT have high similarities and the missing MDST part differs from the MDCT only in its basis functions~\cite{Chen_2010}. Thus, our DNN-based post-filter can infer this missing information in the hidden layers based on the past context and the current MDCT input.

\subsection{Mask Analysis}\label{ss_oracle_expt}
In order to understand the impact of the magnitude mask, an oracle experiment is performed where the ideal magnitude mask computed using Eq.~\eqref{equ_IRM} is applied on the MDCT coefficients as shown in Eq.~\eqref{equ_Mask_mdct}. Since the mask values are unbounded, a threshold $\alpha$ is applied to constrain the mask values to be within [0, $\alpha$]. The bounded mask $\tilde{M}(k,n)$ can be defined as:

\begin{equation}\label{modified_mask}
	\tilde{M}(k,n) =
	\begin{cases}
		M(w,k)       & \quad \text{if} \quad M(w,k) \leq \alpha\\
		\alpha  & \quad \text{if} \quad M(w,k) > \alpha
	\end{cases}.
\end{equation}

\begin{figure}[t]
	\centering
	\includegraphics[width=\linewidth]{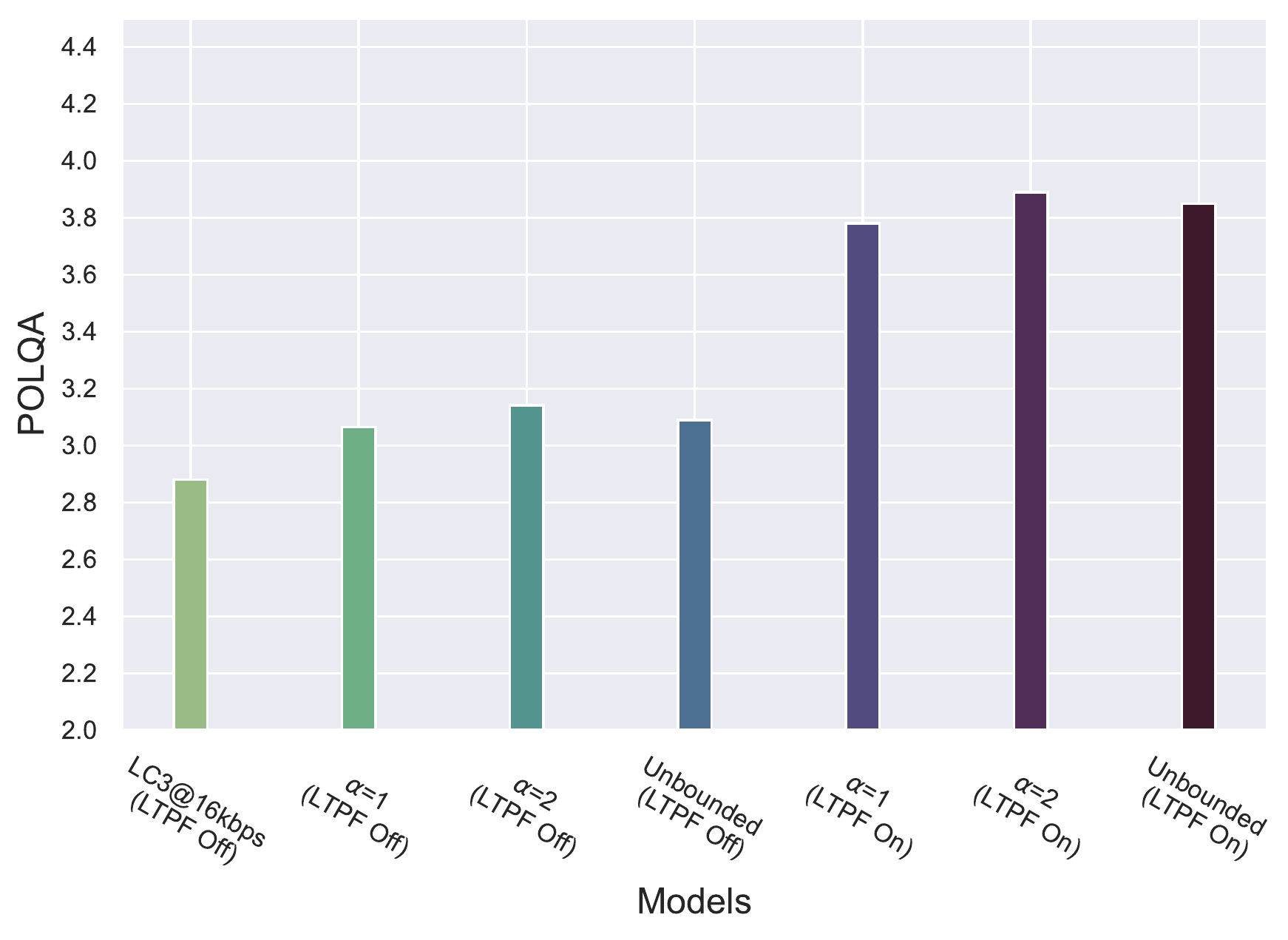}
	\caption{POLQA score evaluation of the performance of the ideal magnitude mask on MDCT with $\alpha$ as upper limit for the mask}
	\label{fig:oracle_exp}
\end{figure}

Fig.~\ref{fig:oracle_exp} compares the average Perceptual Objective Listening Quality Assessment (POLQA)~\cite{POLQA} results of applying the magnitude mask on MDCT coefficients with different $\alpha$ values as an upper bound. It can be observed that the bounding value $\alpha = 2$ can be considered as an ideal upper limit as it provides a similar quality improvement as an unbounded mask. The threshold is important in order to clip the range of values to be predicted, and then ease the DNN task. 

The assessment also validates the usage of mask derived from MCLT magnitude on MDCT coefficients. It shows that mask based post-filter can operate with and without the internal LTPF providing quality improvement over the coded signal in either case. Best performance is observed when the proposed post-filter operates in conjunction with LTPF. 

\section{Experimental Setup}
\subsection{Model}\label{model}
A CNN based encoder-decoder (CED) architecture is implemented as shown in Table~\ref{tab:CNN_architecture} largely inspired from model used in~\cite{Korse_2020}. The input to the DNN is MDCT coefficients of size 160 each for 5 past frames and 1 current frame. Each layer of the CED uses batch normalization and ELU (Exponential Linear Unit) as activation function. Skip connections are used between encoder and decoder layers with required zero-padding inserted to match the \texttt{frequencyBins} dimensions. The output layer uses sigmoid activation function multiplied with a factor 2 in order to estimate the real-valued mask in range [0,2]. The model is trained with the ADAM optimizer~\cite{Kingma2014} with a learning rate of 0.001 and a batch size of 32. Training is done till convergence using early stopping.

\begin{table}[t]
	\setlength{\arrayrulewidth}{0.4mm}
	\centering
	\resizebox{0.47\textwidth}{!}{
		\begin{tabular}{ |c|c|c|c|  }
			\hline
			\textbf{Layer name} & \textbf{Input} & \textbf{Hyperparameter} & \textbf{Output} \\
			\hline
			Reshape & $6 \times 160$ & - & $1 \times 6 \times 160$ \\
			\hline
			Conv2d\textunderscore1 & $1 \times 6 \times 160$ & $2 \times 3$, (1,2), 16  & $16 \times 5 \times 79$ \\
			\hline
			Conv2d\textunderscore2 & $16 \times 5 \times 79$ & $2 \times 3$, (1,2), 32  & $32 \times 4 \times 39 $ \\
			\hline
			Conv2d\textunderscore3 & $32 \times 4 \times 39 $ & $2 \times 3$, (1,2), 64  & $64 \times 3 \times 19 $ \\
			\hline
			Conv2d\textunderscore4 &  $64 \times 3 \times 19 $ & $2 \times 3$, (1,2), 128  & $128 \times 2 \times 9 $ \\
			\hline
			Deconv2d\textunderscore1 & $128 \times 2 \times 9 $ & $2 \times 3$, (1,2), 64  & $64 \times 3 \times 19 $ \\
			\hline
			Deconv2d\textunderscore2 & $128 \times 3 \times 19 $ & $2 \times 3$, (1,2), 32  & $32 \times 4 \times 39 $ \\
			\hline
			Deconv2d\textunderscore3 & $64 \times 4 \times 39 $ & $2 \times 3$, (1,2), 16  & $16 \times 5 \times 79 $ \\
			\hline
			Deconv2d\textunderscore4 & $32 \times 5 \times 79 $ & $2 \times 3$, (1,2), 1  & $1 \times 6 \times 159 $ \\
			\hline
			Conv2d\textunderscore5 & $1 \times 6 \times 160 $ & $6 \times 1$, (1,1), 1  & $1 \times 1 \times 160 $ \\
			\hline
			Flatten & $1 \times 1 \times 160 $ & - & $1 \times 160 $ \\
			\hline
	\end{tabular} }
	\vspace{0.04\linewidth}      
	\caption{Architecture of our proposed CED. The input and output size is given as \texttt{featureMaps} $\times$ \texttt{timesteps} $\times$ \texttt{frequencyBins}. The hyper-parameter is indicated as \texttt{kernelsize, strides, outchannels}. }
	\label{tab:CNN_architecture} 
\end{table}

\subsection{Training and Inference}
\begin{figure}[t]
	\centering
	\includegraphics[width=\linewidth]{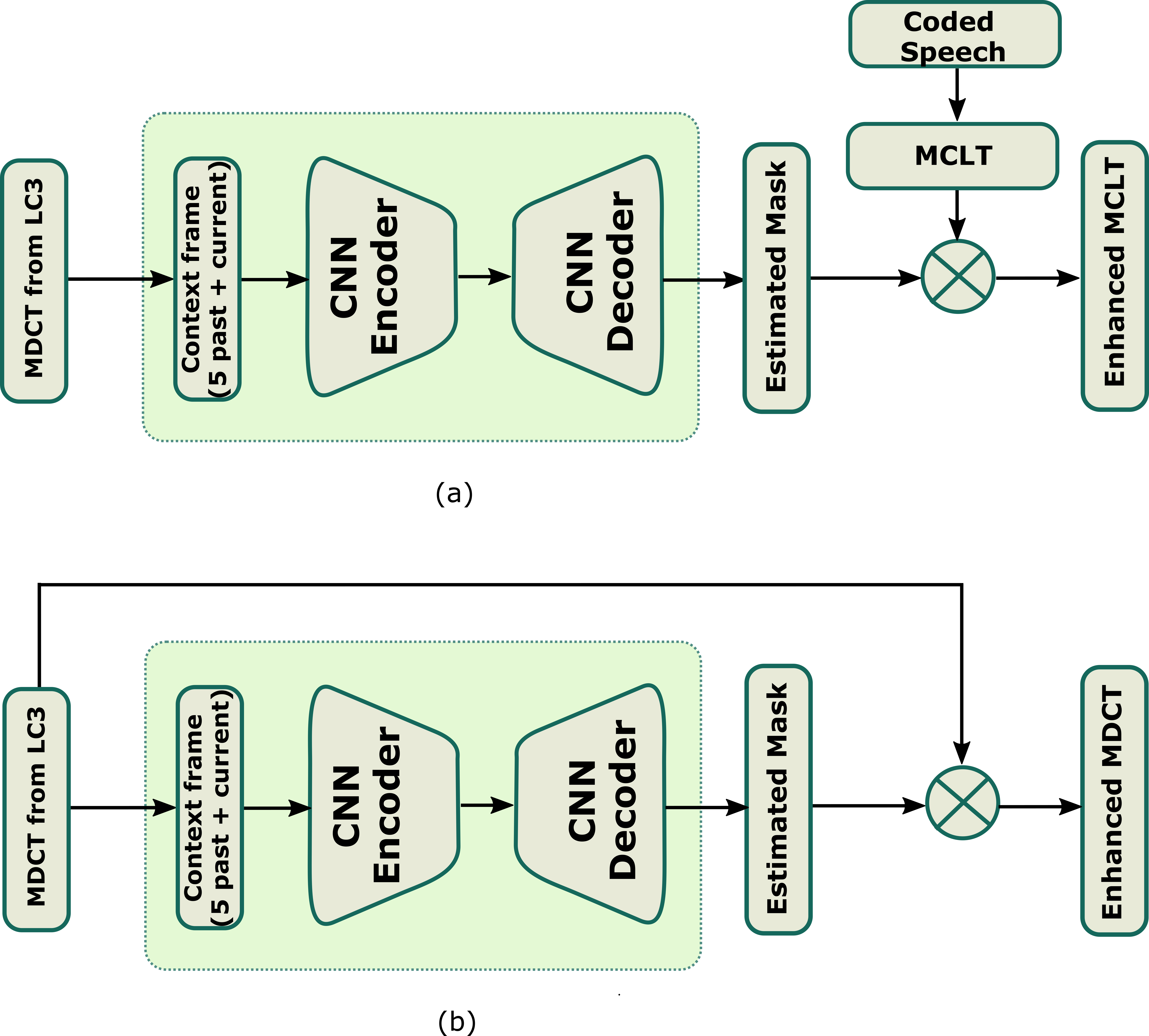}
	\caption{Training and inference phase for MDCT enhancement. Fig. 3a shows the training phase where MDCT is the input to the DNN and MCLT of target and coded speech is used for loss function. Fig. 3b shows the inference phase where input and output are derived from MDCT.}
	\label{fig:training}
\end{figure}

 Based on the analysis shown in~\ref{ss_oracle_expt} which proved the benefits of the magnitude mask applied directly in MDCT domain, we propose the training and inference setup as shown in~Fig~\ref{fig:training}. The input to the model is the logarithm of absolute value of MDCT coefficients obtained from core decoding tools of the LC3 decoder. Since speech signal exhibits temporal dependency, the input to the model contains 5 past frame along with current frame stacked together. The MCLT log-magnitude required for training phase is obtained from coded speech for enhancement and original speech for loss function.  During the training phase, the DNN estimates a magnitude mask which is multiplied to MCLT of coded speech for enhancement. The MSE (Mean Squared Error) between log-magnitude of clean speech MCLT and enhanced MCLT is used as a loss function for training. In the inference, however, the estimated mask is directly applied to the MDCT coefficients thus making the inference completely independent of the complex-valued transform.

\subsection{Datasets}
For both training and testing, files are encoded and decoded with LC3 with internal LTPF enabled or disabled at 16~\unit{kbps}. The training is based on NTT-AT~\cite{nttdb:2012} database containing clean speech stereo signal sampled at 48kHz. It is resampled to 16kHz and a passive mono down-mix is obtained from the stereo files. Out of 3690 files, 3612 files are used for training, 198 files for validation and 150 for testing. The MCLT transform is computed using the same low-delay window employed in LC3 codec~\cite{schnell2021lc3}. For signal with sampling rate of 16kHz, the frame size is 10~\unit{ms} and there is a lookahead of 2.5~\unit{ms}. We use the asymmetric low delay window of LC3 of size 320 samples obtaining 160 MCLT magnitudes per frame. Inputs to the model is normalised by the mean and standard deviation calculated over the entire dataset.

\section{Results}

\begin{figure}[t]
	\centering
	\includegraphics[width=\linewidth]{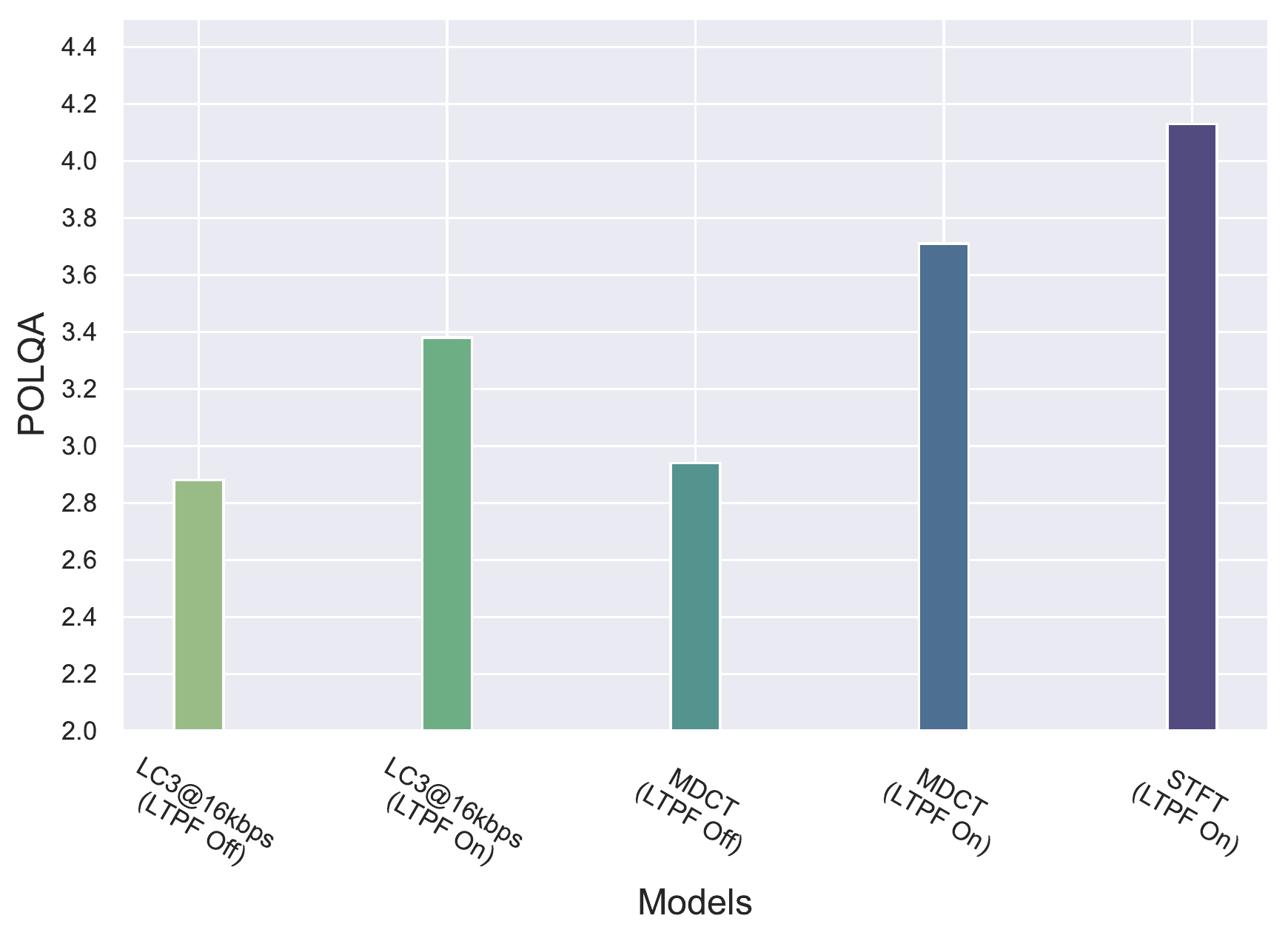}
	\caption{POLQA score evaluation of the performance of our proposed MDCT domain post-filter and its comparison to the LC3 coded speech at 16~\unit{kbps} and baselines.}
	\label{fig_polqa_exp}
\end{figure}

\begin{figure}[t]
	\centering
	\includegraphics[width=\linewidth]{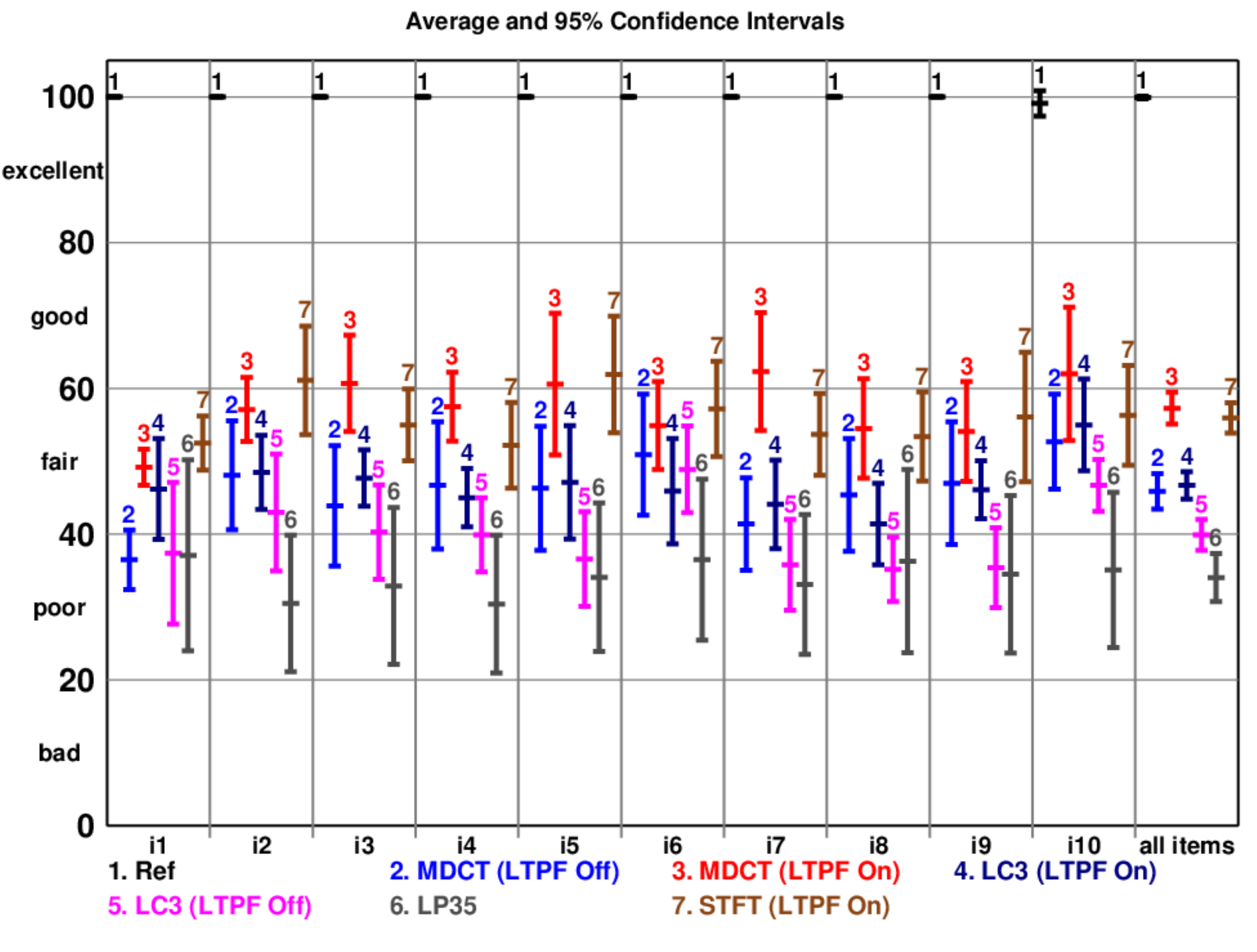}
	\caption{Average MUSHRA scores (Speech) of 10 listeners with Student's-t distribution at 16~\unit{kbps}.}
	\label{fig_MUSHRA_Speech}
\end{figure}


For assessment of the proposed setup, both subjective and objective tests are conducted. For objective assessment, we use POLQA, whereas for subjective assessment, we follow the methodology MUltiple Stimuli with Hidden Reference and Anchor (MUSHRA)~\cite{MUSHRA}. For complete performance evaluation, the assessment is provided for the following configurations:

\begin{itemize}
	\item {} Coded speech with LC3 at 16~\unit{kbps}. Both cases of LTPF enabled (On) and LTPF disabled (Off) at the decoder is analysed.  
	
	\item {} Enhancement of MDCT coefficient from LC3 at 16~\unit{kbps} using proposed post-filter. Both cases of LTPF enabled (On) and LTPF disabled (Off) at the decoder is analysed. No extra delay is introduced over LC3.
	
	\item {}The DNN-based post-filter proposed in~\cite{Korse_2020} is modified to operate up to 8kHz, and works with an forward and inverse STFTs using 32~\unit{ms} frame with 50\% overlap hence operating on 256 frequency bins. This model treats the codec as black-box and takes the decoded time domain signal from codec with LTPF enabled as input. An additional delay of 30~\unit{ms} (3 frames of 10ms each) is then added to coding scheme.
	
\end{itemize}

For comparison of our proposed method, STFT based method serves as baseline for MDCT enhancement with LTPF and coded speech with LTPF serves as baseline for MDCT enhancement without LTPF.

 The POLQA scores are calculated and averaged over 150 test files from NTT-AT database that are not used during training or validation phase. The MUSHRA listening test is done with 10 items in 5 languages taken from various unseen databases thus analysing the robustness and generalization capabilities of our proposed method. Both subjective and objective results confirm that our proposed post-filter improves the perceptually quality of coded speech with and without LTPF. In line with the observation made in the oracle experiment described in the Section~\ref{ss_oracle_expt}, the post-filter along with LTPF provides substantial improvement. The LTPF attenuates the inter harmonic noise at low frequency regions of the spectrum, whereas the DNN based post-filter enhances all regions of the spectrum. Thus, when used in conjunction both the system provides orthogonal improvement leading to better enhancement of the speech signal.

The objective and subjective score differs in their assessment of quality of speech in different configuration. The POLQA score shows that the considered baselines are always better than our proposed post-filter whereas MUSHRA test shows that our post-filter provides good improvement and are comparable in performance to the baselines. From the subjective scores we can infer that our post-filter in MDCT domain is capable of suppressing the quantization noise and generalizes well across different speakers and languages. Moreover, the proposed post-filter does not add any additional delay to the coding scheme and does not require additional frequency transformation unlike the baseline STFT based system. 

In terms of complexity, the proposed post-filter has a complexity of 1.3~\unit{GFLOPS} similar to the STFT based system. Although the complexity of our model is 1000 times more than the heuristic post-filter, we are less than half as complex as other generative models~\cite{Skoglund2019}.

\section{Conclusions}
We proposed a DNN-based post-filter that estimates an optimal magnitude mask derived from an MCLT but applied in the MDCT domain. This method is highly relevant for transform coding, where MDCT is commonly used for its great properties. Integrating the post-filter into the decoder in the MDCT domain eliminates the need for additional algorithmic delay and works directly on quantized coefficients.

Subjective and objective evaluations have demonstrated the effectiveness and robustness of the proposed approach, which can compete with the conventional method of using an additional time-frequency decomposition in a post-processing stage. 

Future work can be devoted to explore the ability of the mask-based approach to enhance signals other than clean speech, such as music and noisy or reverberant speech.



\bibliographystyle{IEEEtran}
\bibliography{refs19}

\begin{thebibliography}{10}
\providecommand{\url}[1]{#1}
\csname url@samestyle\endcsname
\providecommand{\newblock}{\relax}
\providecommand{\bibinfo}[2]{#2}
\providecommand{\BIBentrySTDinterwordspacing}{\spaceskip=0pt\relax}
\providecommand{\BIBentryALTinterwordstretchfactor}{4}
\providecommand{\BIBentryALTinterwordspacing}{\spaceskip=\fontdimen2\font plus
\BIBentryALTinterwordstretchfactor\fontdimen3\font minus
  \fontdimen4\font\relax}
\providecommand{\BIBforeignlanguage}[2]{{%
\expandafter\ifx\csname l@#1\endcsname\relax
\typeout{** WARNING: IEEEtran.bst: No hyphenation pattern has been}%
\typeout{** loaded for the language `#1'. Using the pattern for}%
\typeout{** the default language instead.}%
\else
\language=\csname l@#1\endcsname
\fi
#2}}
\providecommand{\BIBdecl}{\relax}
\BIBdecl

\bibitem{AMRWB+}
\BIBentryALTinterwordspacing
3GPP, ``{Audio codec processing functions; Extended Adaptive Multi-Rate -
  Wideband (AMR-WB+) codec; Transcoding functions},'' {3rd Generation
  Partnership Project (3GPP)}, TS {26.290}, 12. [Online]. Available:
  \url{https://www.3gpp.org/ftp/Specs/archive/}
\BIBentrySTDinterwordspacing

\bibitem{EVS:2014alt}
\BIBentryALTinterwordspacing
------, ``{TS 26.445, EVS Codec Detailed Algorithmic Description; 3GPP
  Technical Specification (Release 12)},'' {3rd Generation Partnership Project
  (3GPP)}, TS {26.445}, 12 2014. [Online]. Available:
  \url{http://www.3gpp.org/ftp/Specs/html-info/26445.htm}
\BIBentrySTDinterwordspacing

\bibitem{LC3:2018Std}
ESTI, ``{TR 103 590: Digital Enhanced Cordless Telecommunications (DECT); Study
  of Super Wideband Codec in DECT for narrowband, wideband and super-wideband
  audio communication including options of low delay audio connections},''
  {European Telecommunications Standards Institute (ETSI)}, TR {103 590}, 2018.

\bibitem{schnell2021lc3}
M.~Schnell, E.~Ravelli, J.~Büthe, M.~Schlegel, A.~Tomasek, A.~Tschekalinskij,
  J.~Svedberg, and M.~Sehlstedt, ``Lc3 and lc3plus: The new audio transmission
  standards for wireless communication,'' in \emph{Audio Engineering Society
  Convention 150}, May 2021.

\bibitem{disch2016intelligent}
\BIBentryALTinterwordspacing
S.~Disch, A.~Niedermeier, C.~R. Helmrich, C.~Neukam, K.~Schmidt, R.~Geiger,
  J.~Lecomte, F.~Ghido, F.~Nagel, and B.~Edler, ``Intelligent gap filling in
  perceptual transform coding of audio,'' in \emph{Audio Engineering Society
  Convention 141}, Sep 2016. [Online]. Available:
  \url{http://www.aes.org/e-lib/browse.cfm?elib=18465}
\BIBentrySTDinterwordspacing

\bibitem{Fuchs_2015}
G.~{Fuchs}, C.~R. {Helmrich}, G.~{Marković}, M.~{Neusinger}, E.~{Ravelli}, and
  T.~{Moriya}, ``Low delay lpc and mdct-based audio coding in the evs codec,''
  in \emph{2015 IEEE International Conference on Acoustics, Speech and Signal
  Processing (ICASSP)}, 2015, pp. 5723--5727.

\bibitem{chen1995adaptive}
{Juin-Hwey Chen} and A.~{Gersho}, ``Adaptive postfiltering for quality
  enhancement of coded speech,'' \emph{IEEE Transactions on Speech and Audio
  Processing}, vol.~3, no.~1, pp. 59--71, Jan 1995.

\bibitem{Das2020_ESSV}
S.~Das and T.~Bäckström, ``Low-complexity postfilter using mdct-domain for
  speech and audio coding,'' in \emph{Studientexte zur Sprachkommunikation:
  Elektronische Sprachsignalverarbeitung 2020}, R.~Böck, I.~Siegert, and
  A.~Wendemuth, Eds.\hskip 1em plus 0.5em minus 0.4em\relax TUDpress, Dresden,
  2020, pp. 109--116.

\bibitem{Korse_2020}
S.~{Korse}, K.~{Gupta}, and G.~{Fuchs}, ``Enhancement of coded speech using a
  mask-based post-filter,'' in \emph{ICASSP 2020 - 2020 IEEE International
  Conference on Acoustics, Speech and Signal Processing (ICASSP)}, 2020, pp.
  6764--6768.

\bibitem{Biswas2020}
A.~{Biswas} and D.~{Jia}, ``Audio codec enhancement with generative adversarial
  networks,'' in \emph{ICASSP 2020 - 2020 IEEE International Conference on
  Acoustics, Speech and Signal Processing (ICASSP)}, 2020, pp. 356--360.

\bibitem{Skoglund2019}
J.~Skoglund and J.-M. Valin, ``Improving opus low bit rate quality with neural
  speech synthesis,'' in \emph{arXiv preprint arXiv:1905.04628}, 2019.

\bibitem{Kuech_2007}
F.~{Kuech} and B.~{Edler}, ``Aliasing reduction for modified discrete cosine
  transform domain filtering and its application to speech enhancement,'' in
  \emph{2007 IEEE Workshop on Applications of Signal Processing to Audio and
  Acoustics}, 2007, pp. 131--134.

\bibitem{Koizumi2018}
Y.~{Koizumi}, N.~{Harada}, Y.~{Haneda}, Y.~{Hioka}, and K.~{Kobayashi},
  ``End-to-end sound source enhancement using deep neural network in the
  modified discrete cosine transform domain,'' in \emph{2018 IEEE International
  Conference on Acoustics, Speech and Signal Processing (ICASSP)}, 2018, pp.
  706--710.

\bibitem{wangmdct}
Y.~Wang, L.~Yaroslavsky, M.~Vilermo, and M.~Vaananen, ``Some peculiar
  properties of the mdct,'' in \emph{WCC 2000 - ICSP 2000. 2000 5th
  International Conference on Signal Processing Proceedings. 16th World
  Computer Congress 2000}, vol.~1, 2000, pp. 61--64 vol.1.

\bibitem{Chen_2010}
S.~Chen, N.~Xiong, J.~Hyuk, M.~Chen, and R.~Hu, ``Spatial parameters for audio
  coding: {MDCT} domain analysis and synthesis,'' \emph{Multimedia Tools
  Appl.}, vol.~48, 06 2010.

\bibitem{POLQA}
\BIBentryALTinterwordspacing
\emph{Perceptual objective listening quality assessment ({POLQA})}, {ITU-T}
  Recommendation {P.863}, 2011. [Online]. Available:
  \url{http://www.itu.int/rec/T-REC-P.863/en}
\BIBentrySTDinterwordspacing

\bibitem{Kingma2014}
D.~Kingma and J.~Ba, ``Adam: A method for stochastic optimization,'' in
  \emph{arXiv preprint arXiv:1412.6980}, 2014.

\bibitem{nttdb:2012}
\BIBentryALTinterwordspacing
NTT-AT, ``Super wideband stereo speech database,''
  http://www.ntt-at.com/product/widebandspeech, accessed: 09.09.2014. [Online].
  Available: \url{http://www.ntt-at.com/product/widebandspeech}
\BIBentrySTDinterwordspacing

\bibitem{MUSHRA}
{Recommendation BS.1534}, \emph{Method for the subjective assessment of
  intermediate quality levels of coding systems}, {ITU-R}, 2003.

\end{thebibliography}

\end{document}